\documentclass[10pt,journal]{IEEEtran}
\usepackage{amsfonts}
\usepackage{amsmath} 
\usepackage{array}
\usepackage{cite}
\usepackage{algorithm}  
\usepackage{algorithmic}

\usepackage{textcomp}
\usepackage{stfloats}
\usepackage{url}
\usepackage{verbatim}
\usepackage{graphicx}
\usepackage{amssymb}
\usepackage{color}
\usepackage{graphics}
\usepackage{float}
\usepackage{caption}
\usepackage{subcaption}
\usepackage{epstopdf}
\usepackage{epsfig}
\usepackage{bbm}
\usepackage[T1]{fontenc}

\newtheorem{remark}{\underline{Remark}}

\usepackage{balance}

\begin{document}

\title{Robust Beamforming Design for Secure Near-Field ISAC Systems}

\author{Ziqiang Chen,~\IEEEmembership{Student Member, IEEE}, Feng Wang,~\IEEEmembership{Member, IEEE}, Guojun Han,~\IEEEmembership{Senior Member, IEEE}, \\ Xin Wang,~\IEEEmembership{Fellow, IEEE}, and Vincent K. N. Lau,~\IEEEmembership{Fellow, IEEE}

\thanks{Z. Chen, F. Wang, and G. Han are with the School of Information Engineering, Guangdong University of Technology, Guangzhou 510006, China (e-mail: czq3215626@163.com, fengwang13@gdut.edu.cn, gjhan@gdut.edu.cn). F. Wang is the corresponding author.}

\thanks{X. Wang is with the School of Information Science and Technology, Fudan University (e-mail: xwang11@fudan.edu.cn).}

\thanks{V. K. N. Lau is with the Department of Electronic and Computer Engineering, The Hong Kong University of Science and Technology, Hong Kong (e-mail: eeknlau@ust.hk).}
\vspace{-0.5cm}

}

\maketitle
\begin{abstract}
This letter investigates the robust beamforming design for a near-field secure integrated sensing and communication (ISAC) system with multiple communication users (CUs) and targets, as well as multiple eavesdroppers. Taking into account the channel uncertainty constraints, we maximize the minimum sensing beampattern gain for targets, subject to the minimum signal-to-interference-plus-noise ratio (SINR) constraint for each CU and the maximum SINR constraint for each eavesdropper, as well as the ISAC transmit power constraint. The formulated design problem is non-convex. As a low-complexity suboptimal solution, we first apply the S-Procedure to convert semi-infinite channel uncertainty constraints into linear matrix inequalities (LMIs) and then use the state-of-the-art sequential rank-one constraint relaxation (SROCR) method to address the rank-one constraints. The numerical results show that the proposed ISAC beamforming design scheme outperforms the existing semidefinite relaxation (SDR) and other baseline schemes, and it significantly enhances security and robustness for near-field ISAC systems.
\end{abstract}


\section{Introduction}
Currently, integrated sensing and communication (ISAC) has emerged as a cutting-edge paradigm that seamlessly integrates communication and sensing functionalities in 6G\cite{r1,r2}. The ISAC system enables the joint optimization of communication and sensing resources, paving the way for new applications in areas such as autonomous driving, smart cities, and artificial internet of things (AIoT) networks.

In contrast to far-field communication, near-field communication uses spherical wavefronts to focus beams at specific locations, introducing new degrees of freedom (DoFs) and research opportunities\cite{r3,r4}. However, near-field ISAC research is limited. The work \cite{r5} investigates a near-field sensing, poisitioning, and communication framework with a novel double-array structure, where a joint angle and distance Cramer-Rao bound is derived and minimized. The work \cite{r6} analyzes near-field unlink/downlink ISAC performance. The work \cite{r7} investigates a full-duplex near-field ISAC system to minimize the transmit power subject to both communication and sensing rate constraints. 

In near-field ISAC systems with the presence of eavesdroppers, countering against information leakage threats in the physical layer is still important to maintain ISAC reliability for practical applications. There are some initial research efforts for secure near-field ISAC systems. For example, taking into account eavesdroppers and beam split effect in near-field wideband MIMO systems, a true-time delayer-based beamfocusing optimization scheme is proposed to maximize the sum secrecy capacity\cite{r8}. For near-field secure communications, hybrid beamforming design solutions are investigated to maximize the secrecy capacity \cite{r9,r10}. 

Note that the aforementioned studies \cite{r5,r6,r7,r8,r9,r10} are all based on the assumption of perfect channel state information (CSI). Robust beamforming design frameworks are proposed in \cite{r11,r12} for reconfigurable intelligent surface (RIS)-based far-field systems under CSI errors. The near-field CSI acquisition and robust beamforming design solutions for near-field secure ISAC systems remain underdeveloped\cite{r13}. Also, during the target detection in practice, existing positioning methods cannot fully obtain near-field target position information\cite{r14}.

In this letter, we propose a robust beamforming design for near-field secure ISAC systems under imperfect CSI, where the multi-antenna base station (BS) simultaneously communicates with multiple single-antenna communication users (CUs) and senses multiple targets, and multiple single-antenna eavesdroppers intend to incept the communication symbols from the BS to the CUs. All the CUs, targets, and eavesdroppers are assumed to be located in the BS's near-field area. By jointly optimizing the communication beamforming vectors for CUs and dedicated sensing covariance matrix for targets, our objective is to maximize the minimum sensing beampattern gain across all targets, subject to the worst-case signal-to-interference-plus-noise ratio (SINR) requirements for each CU, information leakage SINR for each eavesdropper to intercept each CU, and the BS's transmit power constraints. 

In order to address the semi-infinitely many constraints due to norm-bounded CSI errors, we leverage the S-Procedure\cite{r15} with relaxation variables to {\em equivalently} reformulate them into tractable linear matrix inequalities (LMIs). Then, we employ a sequential rank-one constraint relaxation (SROCR) method\cite{r16} to convert the rank-one constraints into the difference of matrix trace and largest eigenvalue. The proposed robust SROCR-based design solution is shown to efficiently converge to a local optimal solution. Extensive numerical results demonstrate the effectiveness of the robust near-field secure ISAC design solution, which achieves enhanced secure ISAC performance compared to far-field counterparts.

\section{System Model and Problem Formulation}

 \begin{figure}
    \centering
    \includegraphics[width=2.0in]{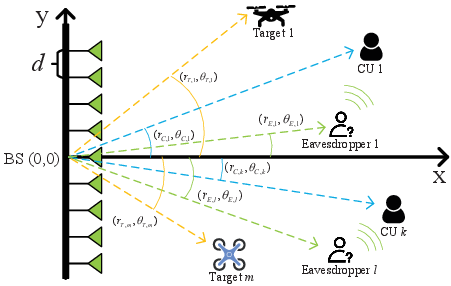}
    \caption{The near-field secure ISAC system model.}\label{fig1}
\end{figure}

 As shown in Fig.~\ref{fig1}, we consider a near-field secure ISAC system model, where the BS is assumed to be equipped with a uniform linear array (ULA) of $N$ antennas and the spacing between each antenna is $d=\lambda/2$. We assume that a set ${\cal K}=\{1,...,K\}$ of single-antenna legitimate CUs and a set ${\cal M}=\{1,...,M\}$ of targets of interest to be sensed, as well as a set ${\cal L}=\{1,...,L\}$ of single-antenna eavesdroppers, are randomly located in the near-field area of the BS\footnote{Note that instead of treating the eavesdroppers as targets to be sensed, this work assumes that these eavesdroppers' locations have already been acquired e.g., via target sensing methods \cite{r8,r9,r10}.}. Accordingly, the distance (e.g., $r$) between the BS and each CU/target/eavesdropper is smaller than the Rayleigh distance $d_R=2D^2/\lambda$ but larger than the Fresnel distance $d_F=0.5\sqrt{D^3/\lambda}$\cite{r4}, i.e., $d_F\leq r\leq d_R$, where $D=(N-1)d$ denotes the BS antenna array aperture and $\lambda$ denotes the wavelength of the electromagnetic signal.

 \subsection{Near-Field Channel Model}
 We assume the center of the BS's ULA is located at the origin of the coordinate system. Then, the coordinate of the $n$-th antenna element of the BS is given by $(0,\frac{-N-1+2n}{2}d)$, where $n=1,...,N$. Denote by $(r,\theta)$ the polar coordinate of one point in the near-field area of BS. Then, the BS's near-field array steering vector towards the point $(r,\theta)$ for line-of-sight (LoS) communication is modeled as $\boldsymbol{\alpha}(r, \theta)=[e^{-j \frac{2 \pi}{\lambda}r_{1}},..., e^{-j \frac{2 \pi}{\lambda} r_{N}}]^{T}$, where $r_n=\sqrt{r^2+\frac{(-N-1+2n)^2d^2}{4}+(N+1-2n)rd\sin\theta}$ denotes the distance between the $n$-th antenna element of the BS and the point $(r,\theta)$. Denote by $(r_{C,k}, \theta_{C,k})$, $(r_{T,m}, \theta_{T,m})$, and $(r_{E,l}, \theta_{E,l})$ the polar coordinates of CU $k\in{\cal K}$, target $m\in{\cal M}$, and eavesdropper $l\in{\cal L}$, respectively. The near-field channel vectors from the BS to CU $k$, target $m$, and eavesdropper $l$ are respectively modeled as\footnote{Note that we focus on the mmWave near-field ISAC scenarios where the LoS component typically dominates the non-LoS (NLoS) components, and the NLoS components can thus be safely negligible. By incorporating NLoS components in \eqref{eq.LoS}, the proposed robust beamforming design scheme can be extended to the LoS/NLoS near-field ISAC scenarios.}
\begin{subequations}\label{eq.LoS}
\begin{align}
    &\boldsymbol{h}_{C,k}=h^{\text{LoS}}_{C,k} \boldsymbol{\alpha}(r_{C,k},\theta_{C,k})\\
    &\boldsymbol{h}_{T,m}=h^{\text{LoS}}_{T,m} \boldsymbol{\alpha}(r_{T,m},\theta_{T,m})\\
    &\boldsymbol{h}_{E,l}=h^{\text{LoS}}_{E,l} \boldsymbol{\alpha}(r_{E,l},\theta_{E,l}),
\end{align}
\end{subequations}
where $h^{\text{LoS}}_{C,k}=\sqrt{\rho_0 r_{C,k}^{-\alpha}}$, $h^{\text{LoS}}_{T,m}=\sqrt{\rho_0 r_{T,m}^{-\alpha}}$, and $h^{\text{LoS}}_{E,l}=\sqrt{\rho_0 r_{E,l}^{-\alpha}}$ denote the large-scale LoS channel pathloss from the BS to CU $k$, target $m$, and eavesdropper $l$, respectively, $\rho_0=\frac{\lambda}{4\pi}$ denotes the reference pathloss at a distance of one meter, and $\alpha$ denotes the pathloss exponential number. 

Note that due to the channel estimation errors from the BS to multiple CUs/targets/eavesdroppers, the perfect CSI $\{\boldsymbol{h}_{C,k},\boldsymbol{h}_{T,m},\boldsymbol{h}_{E,l}\}$ is generally not available at the BS. As widely adopted in ISAC design works\cite{r11}, we consider a bounded CSI error model, where the CSI errors are subject to deterministic norm constraints. Therefore, we adopt an additive CSI error model as
$\boldsymbol{h}_{C,k} = \hat{\boldsymbol{h}}_{C,k} + \Delta_{\boldsymbol{h}_{C,k}}$, $\boldsymbol{h}_{T,m} = \hat{\boldsymbol{h}}_{T,m} + \Delta_{\boldsymbol{h}_{T,m}}$, $\boldsymbol{h}_{E,l} = \hat{\boldsymbol{h}}_{E,l} + \Delta_{\boldsymbol{h}_{E,l}}$,
where $\|\Delta_{\boldsymbol{h}_{C,k}}\|\leq \epsilon_{C,k}$, $\|\Delta_{\boldsymbol{h}_{T,m}}\| \leq \epsilon_{T,m}$, $\|\Delta_{\boldsymbol{h}_{E,l}}\| \leq \epsilon_{E,l}$; $\hat{\boldsymbol{h}}_{C,k}$ ($\hat{\boldsymbol{h}}_{T,m}$, $\hat{\boldsymbol{h}}_{E,l}$) and $\Delta_{\boldsymbol{h}_{C,k}}$ ($\Delta_{\boldsymbol{h}_{T,m}}$, $\Delta_{\boldsymbol{h}_{E,l}}$) denote the channel estimate and error for CU $k$ (target $m$, eavesdropper $l$), respectively; $\epsilon_{C,k}$, $\epsilon_{T,m}$, and $\epsilon_{E,l}$ denote the norm bounds of channel error for CU $k$, target $m$, and eavesdropper $l$, respectively. Note that it remains challenging to obtain the CSI $\{\hat{\boldsymbol{h}}_{C,k},\hat{\boldsymbol{h}}_{T,m},\hat{\boldsymbol{h}}_{E,l}\}$ via angle and distance based beam training methods\cite{r13}. By extending the two-stage near-field beam training method in\cite{r13}, it is worth further investigation for joint CSI acquisition and robust beamforming ISAC designs in near-field secure ISAC systems.

\subsection{Signal Model}
Denote by $s_k\in\mathbb{C}$ the information signal for each CU $k\in{\cal K}$, which are assumed to be independently and identically distributed (i.i.d.) with unit power, i.e., $\mathbb{E}[|{s}_k|^2 ]=1$, $\forall k \in \mathcal{K}$. Denote by $\boldsymbol{s}_0\in\mathbb{C}^{N\times1}$ the dedicated sensing symbol vector, and let $\boldsymbol{R}_0=\mathbb{E}[\boldsymbol{s}_0 \boldsymbol{s}_0^{H}]\in\mathbb{C}^{N\times N}$ denote the BS sensing covariance matrix. The ISAC transmit signal of the BS is thus modeled as $\boldsymbol{x}=\sum_{k=1}^K \boldsymbol{w}_k {s}_k+\boldsymbol{s}_0
$, where $\boldsymbol{w}_k \in \mathbb{C}^{N \times 1}$ denotes the transmit beamforming vector for CU $k$. Consequently, the equivalent baseband received signal at each CU $k\in{\cal K}$ and each eavesdropper $l\in{\cal L}$ are respectively modeled as
 \begin{subequations}\label{eq.def1}
\begin{align}
    &y_{C,k}  =\boldsymbol{h}_{C,k}^{H} \boldsymbol{w}_k {s}_k+\sum_{i \neq k} \boldsymbol{h}_{C,k}^{H} \boldsymbol{w}_i {s}_i+\boldsymbol{h}_{C,k}^{H} \boldsymbol{s}_0+n_{C,k} \\
   & {y}_{E,l}  =\sum_{i=1}^K \boldsymbol{h}_{E,l}^{H} \boldsymbol{w}_i {s}_i+\boldsymbol{h}_{E,l}^{H} \boldsymbol{s}_0+n_{E,l},
\end{align}
\end{subequations}
 where $n_{C,k} \sim \mathcal{C N}(0, \sigma_{C,k}^2)$ and $n_{E,l} \sim \mathcal{C N}(0, \sigma_{E,l}^2)$ denote the additive white Gaussian noise (AWGN) at CU $k$ and eavesdropper $l$, respectively. 
 
 Based on (\ref{eq.def1}a), the worst-case SINR of CU $ k\in\mathcal{K}$ to decode its intended symbol $s_k$ is given by
  \begin{align}\label{eq.SINR_Ck}
&\gamma_{C,k}=\min_{\|\Delta_{\boldsymbol{h}_{C,k}}\| \leq \epsilon_{C,k}} \frac{|\boldsymbol{h}_{C,k}^{H} \boldsymbol{w}_k|^2}{\sum_{i \neq k}^K|\boldsymbol{h}_{C,k}^{H} \boldsymbol{w}_i|^2+|\boldsymbol{h}_{C,k}^{H} \boldsymbol{s}_0|^2+\sigma_{C,k}^2},\notag \\
&~~~~~~~~\qquad\qquad\qquad\qquad\qquad\qquad\qquad\qquad\forall k \in \mathcal{K}.
 \end{align}
 
Based on (\ref{eq.def1}b), the maximal SINR for eavesdropper $l$ to intercept CU $k$’s symbol $s_k$ is given by 
 \begin{align} \label{eq.SINR_Elk}
&\gamma_{E,l,k} = \max_{\|\Delta_{\boldsymbol{h}_{E,l}}\| \leq \epsilon_{E,l}} \frac{|\boldsymbol{h}_{E,l}^{H} \boldsymbol{w}_k|^2}{\sum_{i\neq k}|\boldsymbol{h}_{E,l}^{H} \boldsymbol{w}_i|^2+|\boldsymbol{h}_{E,l}^{H} \boldsymbol{s}_0|^2+\sigma_{E,l}^2},\notag \\
&~~~~~~~~\qquad\qquad\qquad\qquad\qquad\qquad\forall k \in \mathcal{K}, l \in \mathcal{L}.
 \end{align}
  
 Note that the sensing beampattern gain (i.e., the sensing illumination power) for each target of interest is adopted as a key sensing performance metric, and a larger sensing beampattern gain typically leads to a better sensing capability for obtaining the target's properties (e.g., position and size). By taking into account the CSI errors, the worst-case sensing beampattern gain for target $m\in{\cal M}$ is then modeled as
 \begin{align}
\rho_{T,m} = \min_{\|\Delta_{\boldsymbol{h}_{T,m}}\|\leq \epsilon_{T,m}} \boldsymbol{h}_{T,m}^{H}\Big(\sum_{k=1}^K \boldsymbol{w}_k \boldsymbol{w}_k^H+\boldsymbol{R}_0\Big) \boldsymbol{h}_{T,m}.
 \end{align}

 \subsection{Problem Formulation}
In this letter, we are interested in addressing the optimization problem by taking into account CSI errors. The goal is to maximize the minimum worst-case sensing beampattern gain among $M$ targets, by jointly optimizing the communication beamforming vectors $\{\boldsymbol{w}_k\}_{k \in \mathcal{K}}$ and the sensing covariance matrix $\boldsymbol{R}_0$. Therefore, by introducing an auxiliary variable $t$ as the lower bound of the sensing beampattern gain, the near-field robust secure ISAC design problem is formulated as
\begin{subequations}\label{eq.p1}
\begin{align}
 &\text { (P1): } \max _{\{\boldsymbol{w}_k\}_{k \in \mathcal{K}}, \boldsymbol{R}_0\succeq 0,t} t \\
 & \text { s.t. } ~~\sum_{k=1}^K\|\boldsymbol{w}_k\|^2+\operatorname{tr}\left(\boldsymbol{R}_0\right) \leq P_0
 \\
&~~~~~~\rho_{T,m}\ge t,\forall m\in{\cal M} \\
 &~~~~~~ \gamma_{C,k} \geq \bar{\gamma}_{C,k}, \forall k \in \mathcal{K}\\
 &~~~~~~\gamma_{E,l,k} \leq \bar{\gamma}_{E,l,k}, \forall l \in \mathcal{L}, \forall k \in \mathcal{K},
\end{align}
\end{subequations}
 where $P_0$ denotes the BS transmission power budget, $\bar{\gamma}_{C,k}$ and $\bar{\gamma}_{E,l,k}$ denote the minimum SINR threshold for CU $k$ and the maximum tolerable information leakage SINR for eavesdropper $l$ to intercept CU $k$, respectively. Note that due to the target sensing channel error norm constraint $\|\Delta_{\boldsymbol{h}_{T,m}}\|\leq\epsilon_{T,m}$, the constraints in (\ref{eq.p1}c)  involve a number of semi-infinite expressions. Also, due to the existence of CSI errors, both (\ref{eq.p1}d) and (\ref{eq.p1}e) involve a number of semi-infinite non-convex SINR constraints. Therefore, problem (P1) is a non-convex problem and challenging to solve.

 \section{Proposed Solution for (P1)}
 In this section, we first apply the S-Procedure to transform the semi-infinitely many constraints into tractable LMIs, and then present a SROCR-based design solution for (P1). 

\subsection{S-Procedure based Problem Transformation}
 To start with, we define $\boldsymbol{W}_k\triangleq \boldsymbol{w}_k \boldsymbol{w}_k^{H}$, $\forall k\in{\cal K}$, and it follows that $\boldsymbol{W}_k\succeq 0$ and $\operatorname{rank}\left(\boldsymbol{W}_k\right) = 1$, $\forall k \in \mathcal{K}$. Furthermore, we define $\boldsymbol{\Psi} \triangleq \sum_{k=1}^K\boldsymbol{W}_k +\boldsymbol{R}_0$. By replacing $\boldsymbol{h}_{T,m}$ with $\hat{\boldsymbol{h}}_{T,m}+\Delta_{\boldsymbol{h}_{T,m}}$, the constraints in (\ref{eq.p1}c) become
  \begin{align} \label{eq.PTm}
     \Delta_{\boldsymbol{h}_{T,m}}^{H} \boldsymbol{\Psi} \Delta_{\boldsymbol{h}_{T,m}}+2 \text{Re}\{\hat{\boldsymbol{h}}_{T,m}^{H} \boldsymbol{\Psi} \Delta_{\boldsymbol{h}_{T,m}}\}+\hat{\boldsymbol{h}}_{T,m}^{H} \boldsymbol{\Psi} \hat{\boldsymbol{h}}_{T,m} \geq t,
 \end{align}
 where $\|\Delta_{\boldsymbol{h}_{T,m}}\| \leq \epsilon_{T,m}$, $\forall m\in{\cal M}$. By using the well-known S-Procedure\cite{r15}, we can equivalently transform \eqref{eq.PTm} into the LMIs constraint as
 \begin{equation}\label{eq.Tm}
\left[\begin{array}{cc}
\mu_{T,m} \boldsymbol{I}_N+\boldsymbol{\Psi} & \boldsymbol{\Psi}^{H}\hat{\boldsymbol{h}}_{T,m} \\
\hat{\boldsymbol{h}}_{T,m}^{H} \boldsymbol{\Psi} & \hat{\boldsymbol{h}}_{T,m}^{H} \boldsymbol{\Psi}\hat{\boldsymbol{h}}_{T,m}-t-\mu_{T,m} \epsilon_{T,m}^2
\end{array}\right] \succeq \mathbf{0},
\end{equation}
where $m\in{\cal M}$ and $\mu_{T,m}\geq 0$ is termed as a slack non-negative variable to be optimized later.

Next, we define $\boldsymbol{\Psi}_{k}\triangleq \frac{1}{\bar{\gamma}_{C,k}}\boldsymbol{W}_k - {\sum_{i=1, i\neq k}^K } \boldsymbol{W}_i - \boldsymbol{R}_0$ and $\boldsymbol{\Psi}_{l,k}\triangleq \frac{1}{\bar{\gamma}_{E,l,k}}\boldsymbol{W}_k- {\sum_{i=1, i\neq k}^K } \boldsymbol{W}_i- \boldsymbol{R}_0$, where $k\in{\cal K}$ and $l\in{\cal L}$. By substituting $\boldsymbol{h}_{C,k}  =\hat{\boldsymbol{h}}_{C,k}+\Delta _{\boldsymbol{h}_{C,k}}$ and $\boldsymbol{h}_{E,l} = \hat{\boldsymbol{h}}_{E,l} + \Delta_{\boldsymbol{h}_{E,l}}$ into  (\ref{eq.p1}d) and (\ref{eq.p1}e), the SINR constraints in (\ref{eq.p1}d) and (\ref{eq.p1}e) can be respectively reformulated as
\begin{subequations}\label{eq.SINR}
\begin{align}
 &\Delta _{\boldsymbol{h}_{C,k}}^H \boldsymbol{\Psi}_{k} \Delta _{\boldsymbol{h}_{C,k}} +2 \operatorname{Re}\{\hat{\boldsymbol{h}}_{C,k}^{H} \boldsymbol{\Psi}_{k} \Delta _{\boldsymbol{h}_{C,k}}\}+ \hat{\boldsymbol{h}}_{C,k}^{H} \boldsymbol{\Psi}_{k} \hat{\boldsymbol{h}}_{C,k}  \nonumber \\
 &\qquad\qquad\qquad\qquad\qquad\qquad\qquad\qquad- \sigma_{C,k}^2 \geq 0 \label{eq:12a} \\
 &\Delta _{\boldsymbol{h}_{E,l}}^{H} \boldsymbol{\Psi}_{l,k} \Delta_{\boldsymbol{h}_{E,l}} +2 \operatorname{Re}\{\hat{\boldsymbol{h}}_{E,l}^{H} \boldsymbol{\Psi}_{l,k} \Delta _{\boldsymbol{h}_{E,l}}\} + \hat{\boldsymbol{h}}_{E,l}^{H} \boldsymbol{\Psi}_{l,k} \hat{\boldsymbol{h}}_{E,l}\nonumber \\
 &\qquad\qquad\qquad\qquad\qquad\qquad\qquad\qquad  - \sigma_{E,l}^2 \leq 0, \label{eq:12b}
\end{align}
\end{subequations}
where $\|\Delta_{\boldsymbol{h}_{C,k}}\| \leq \epsilon_{C,k}$ and $\|\Delta_{\boldsymbol{h}_{E,l}}\| \leq \epsilon_{E,l}$, $\forall l\in{\cal L}$, $k\in{\cal K}$.

Based on the S-Procedure\cite{r15}, we respectively obtain the equivalent LMIs representations of (\ref{eq.SINR}a)  and (\ref{eq.SINR}b) as
\begin{subequations}\label{eq.SINR_S}
\begin{align}
 &\left[\begin{array}{cc}
 \mu_{C,k} \boldsymbol{I}_{N}+\boldsymbol{\Psi}_{k} & \boldsymbol{\Psi}_{k}^{H} \hat{\boldsymbol{h}}_{C,k} \\
 \hat{\boldsymbol{h}}_{C,k}^{H} \boldsymbol{\Psi}_{k} & \hat{\boldsymbol{h}}_{C,k}^{H} \boldsymbol{\Psi}_{k} \hat{\boldsymbol{h}}_{C,k}-\sigma_{C,k}^{2}-\mu_{C,k} \epsilon_{C,k}^{2}
\end{array}\right] \notag \\
&~~~~~~\qquad\qquad\qquad\qquad\qquad\qquad\qquad\qquad\succeq \mathbf{0} \\
 &\left[\begin{array}{cc}
 \mu_{E,l,k} \boldsymbol{I}_N-\boldsymbol{\Psi}_{l,k} & -\boldsymbol{\Psi}_{l,k}^{H}\hat{\boldsymbol{h}}_{E,l} \\
 -\hat{\boldsymbol{h}}_{E,l}^{H} \boldsymbol{\Psi}_{l,k} & \sigma_{E,l}^{2}-\hat{\boldsymbol{h}}_{E,l}^{H} \boldsymbol{\Psi}_{l,k}\hat{\boldsymbol{h}}_{E,l}-\mu_{E,l,k} \epsilon_{E,l}^{2}
 \end{array}\right] \notag \\
 &~~~~~\qquad\qquad\qquad\qquad\qquad\qquad\qquad \qquad\succeq \mathbf{0}, 
\end{align}
\end{subequations}
where both $\mu_{C,k}\geq 0$ and $\mu_{E,l,k}\geq 0$ denote the non-negative slack variables to be optimized later. 

Note that the CSI errors $\{\Delta_{\boldsymbol{h}_{C,k}},\Delta_{\boldsymbol{h}_{T,m}},\Delta_{\boldsymbol{h}_{E,l}}\}$ are subject to norm constraints, which are convex and have non-empty interior points. Based on Theorem 3.3 in \cite{r15}, the equivalence of S-procedure based transformation of \eqref{eq.Tm} and \eqref{eq.SINR_S} is thus guaranteed. With the LMIs constraints \eqref{eq.Tm} and \eqref{eq.SINR_S} at hand, we are ready to equivalently recast problem (P1) as 
\begin{subequations}\label{eq.p2}
 \begin{align}
  &\text { (P2): } \max _{\{\boldsymbol{W}_k\succeq 0\}, \boldsymbol{R}_0\succeq 0, \boldsymbol{\mu}\ge 0, t} ~~  t\\
 &~~~~~~~~~~\text {s.t.} \sum_{k=1}^K \operatorname{tr}(\boldsymbol{W}_k)+\operatorname{tr}(\boldsymbol{R}_0) \leq P_0\\
 &~~~~~~~~~~~~~~~(\ref{eq.Tm})~(\ref{eq.SINR_S}a)~(\ref{eq.SINR_S}b)\\
 &~~~~~~~~~~~~~\operatorname{rank}(\boldsymbol{W}_k) = 1,~\forall k \in \mathcal{K},
 \end{align}
\end{subequations}
where $\boldsymbol{\mu} \triangleq[\mu_{C,1},...,\mu_{C,K}, \mu_{E,1, 1},...,\mu_{E,L,K},\mu_{T, 1},...,\mu_{T,M}]^T$ and $\boldsymbol{\mu}\geq 0$ denotes the element-wise inequality. Due to rank-one constraints (\ref{eq.p2}d), (P2) is a non-convex problem.

\subsection{SROCR-based Solution for (P2)}
 To address the rank-one constraints in (P2), different from dropping the rank-one constraints in the semidefinite relaxation (SDR) method, we employ the state-of-the-art sequential rank-one constraint relaxation (SROCR) algorithm\cite{r16} for (P2) (and equivalent for (P1)).
 
 Specifically, the rank-one constraints in (\ref{eq.p2}d) are equivalently recast as $\lambda_{\max}(\boldsymbol{W}_k)=\operatorname{tr}(\boldsymbol{W}_k)$, $\forall k\in{\cal K}$, where $\lambda_{\max }(\boldsymbol{A})$ and $\operatorname{tr}(\boldsymbol{A})$ denote the largest eigenvalue and the trace of matrix $\boldsymbol{A}$, respectively. Based on the basic idea of SROCR algorithm \cite{r16}, we introduce a relaxation parameter $v_k^{(i-1)}\in[0,1]$ for each rank-one constraint $\operatorname{rank}(\boldsymbol{W}_k)=1$ during the $i$-th iteration. Then, we obtain $\boldsymbol{u}_{k,\max }^{(i-1)H} \boldsymbol{W}_k \boldsymbol{u}_{k,\max }^{(i-1)}\geq v_k^{(i-1)} \operatorname{tr}(\boldsymbol{W}_k)$, $\forall k\in{\cal K}$, where $\boldsymbol{u}_{k,\max }^{(i-1)}$ denotes the eigenvector associated with the largest eigenvalue of matrix $\boldsymbol{W}_k^{(i-1)}$, and $\boldsymbol{W}^{(i-1)}_k$ is obtained during the $(i-1)$-th iteration. 

 In order to implement the SROCR algorithm for (P2), during the $i$-th iteration, we need to solve the following semi-definite program (SDP) problem as 
\begin{subequations}\label{eq.p3}
 \begin{align}
  &\text { (P3): } \max _{\left\{\boldsymbol{W}_k\succeq 0\right\}, \boldsymbol{R}_0 \succeq 0, \boldsymbol{\mu}\geq 0, t}~~ t\\
 &~~~\text { s.t. }  (\ref{eq.p2}b)- (\ref{eq.p2}c)\\
&~~~~~~~~\boldsymbol{u}_{k,\max }^{(i-1)H} \boldsymbol{W}_k \boldsymbol{u}_{k,\max }^{(i-1)}\geq v_k^{(i-1)} \operatorname{tr}(\boldsymbol{W}_k),
\forall k\in{\cal K},
\end{align}
\end{subequations}
which is convex and can thus be efficiently and optimally solved by off-the-shelf convex solvers. 

\begin{algorithm}
\caption{Proposed SROCR-based Solution for (P2)}
    \begin{algorithmic}[1]
    \STATE \textbf{Initialize:} set $i=1$, initialize $\{v_k^{(0)},\delta_k^{(0)}\}$, and set error tolerance level $\epsilon$; obtain the eigenvector $\boldsymbol{u}^{(0)}_{k,\max}$ associated with the largest eigenvalue of $\boldsymbol{W}^{(0)}_k$, $\forall k\in{\cal K}$, where $\{\boldsymbol{W}^{(0)}_k\}$ are obtained by solving the rank-relaxed (P2) (i.e., removing rank-one constraints (\ref{eq.p2}d));    
     \REPEAT
      \STATE \textbf{Check}: if (P3) is feasible, then solve (P3) to obtain $\boldsymbol{W}_k^{(i)}$ and set $\delta_k^{(i)}=\delta_k^{(i-1)}$, $\forall k\in{\cal K}$; otherwise, then set $\delta_k^{(i)}=\frac{1}{2}\delta_k^{(i-1)}$, $\forall k\in{\cal K}$;
     \STATE \textbf{Update}:
      $v_k^{(i)}=\min \left(1, \frac{\lambda_{\max }\left(\boldsymbol{W}_k^{(i)}\right)}{\operatorname{tr}\left(\boldsymbol{W}_k^{(i)}\right)}+\delta_k^{(i)}\right)$, $\forall k\in{\cal K}$;
    \STATE \textbf{Set}: $i\gets i+1$;
       \UNTIL $v_k^{(i)}=1$, $\forall k\in{\cal K}$, and $|t^{(i)}-t^{(i-1)}|\leq \epsilon$.
    \end{algorithmic}
\end{algorithm}

\begin{remark}
 Note that in the proposed SROCR-based Algorithm~1 for (P2), the relaxation parameter $v_k^{(i)}$ is used to control the ratio of the largest eigenvalue to the trace of $\boldsymbol{W}^{(i+1)}_k$ during the $(i+1)$-th iteration, and the value $v_k^{(i)}$ gradually increases from zero to one. By iteratively increasing the value $v_k^{(i)}$, the feasible set of matrix $\boldsymbol{W}_k$ shrinks and the trace of $\boldsymbol{W}_k$ approaches one; when $v_k^{(i)}=1$, we have $\text{rank}(\boldsymbol{W}_k)=1$. Based on Theorem 1 in \cite{r16}, Algorithm~1 is guaranteed to converge to a local (but not necessarily global) Karush-Kuhn-Tucker (KKT) stationary point of (P2) (equivalent to (P1)). The convergence rate of Algorithm~1 depends on the setting of step-size $\{\delta_k^{(0)}\}$ and the update policy of $\{\delta_k^{(i)}\}$ per iteration. Also, a small increase in $\{v_k^{(i)},\delta_k^{(i)}\}$ leads to a slow convergence, but improves the probability of achieving a large objective value~\cite{r16}.
\end{remark}

\begin{remark}
 Note that the computational complexity of solving an SDP (ignoring complexity of linear constraints) is given as $
 \mathcal{O}((\sum_{i=1}^J b_i)^{\frac{1}{2}}(n^3+n^2\sum_{i=1}^J b^2_i+n\sum_{i=1}^J b^3_i)$\cite{r11}, where $n$ denotes the variable number, $J$ is the number of LMIs of size $b_i$. Since problem (P3) involves $(M+K+LK)$ LMIs of size $(N+1)$, $(K+1)$ LMIs of size $N$, and a number of $K(N^2+L+1)+N^2+M$ variables, the computational complexity of the proposed Algorithm~1 is then $
 \mathcal{O}(J_1^{\frac{1}{2}}(n^3+n^2J_2+nJ_3)Q)$, where $n=K(N^2+L+1)+N^2+M$, $J_1=(N+1)(M+K+LK)+N(K+1)$, $J_2=(N+1)^2(M+K+LK)+N^2(K+1)$, $J_3=(N+1)^3(M+K+LK)+N^3(K+1)$, and $Q$ denotes the iteration number of Algorithm~1.
\end{remark}

\section{Numerical Results}
In this section, we provide numerical results to evaluate the proposed robust secure ISAC beamforming design based on the proposed SROCR-based Algorithm~1, where we initialize $v_k^{(0)}=0$, $\delta_k^{(0)}=0.1$, $\forall k\in{\cal K}$, and $\epsilon=10^{-4}$. Each legitimate CU $k\in{\cal K}$ and eavesdropper $l \in{\cal L}$ are randomly distributed within the near-field region of the BS. Unless otherwise specified, we set $N=64$ and BS operating at a frequency of 28~GHz, $K=4$, $L=2$, $M=2$, where the sensing targets are located at $(10\mathrm{~m},-\frac{\pi}{4})$ and $(10\mathrm{~m}, \frac{\pi}{5})$, respectively. The pathloss exponential value is set as $\alpha=2$\cite{r5}. The minimum SINR threshold for CU $k$ is set as $\bar{\gamma}_{C,k}=5$~dB, and the SINR threshold to prevent the eavesdropper $l$ from intercepting the information of CU $k$ is set as $\bar\gamma_{E,l,k}=-3$~dB. The receiver noise variances for all CUs and eavesdroppers are set as $\sigma_{C,k}^2=\sigma_{E,l}^2=-80$~dBm, $\forall k\in{\cal K}$, $l\in{\cal L}$. The transmission power is set as $P_0=30$~dBm. For the CSI error model\cite{r11}, we define $\eta_{C,k}=\frac{\epsilon_{C,k}}{\|\hat{\boldsymbol{h}}_{C,k}\|}$, $\eta_{E,l}=\frac{\epsilon_{E,l}}{\|\hat{\boldsymbol{h}}_{E,l}\|}$ and $\eta_{T,m}=\frac{\epsilon_{T,m}}{\left\|\hat{\boldsymbol{h}}_{T,m}\right\|}$ as the normalized CSI error bounds, Unless otherwise specified, we set $\eta_{C,k}=\eta_{E,l}=\eta_{T,m}=\eta$, $\forall k \in \mathcal{K}$, $l \in \mathcal{L}$, $m \in \mathcal{M}$. 

\begin{figure}
    \centering
    \begin{subfigure}[b]{0.81\linewidth} 
        \centering
        \includegraphics[width=\textwidth]{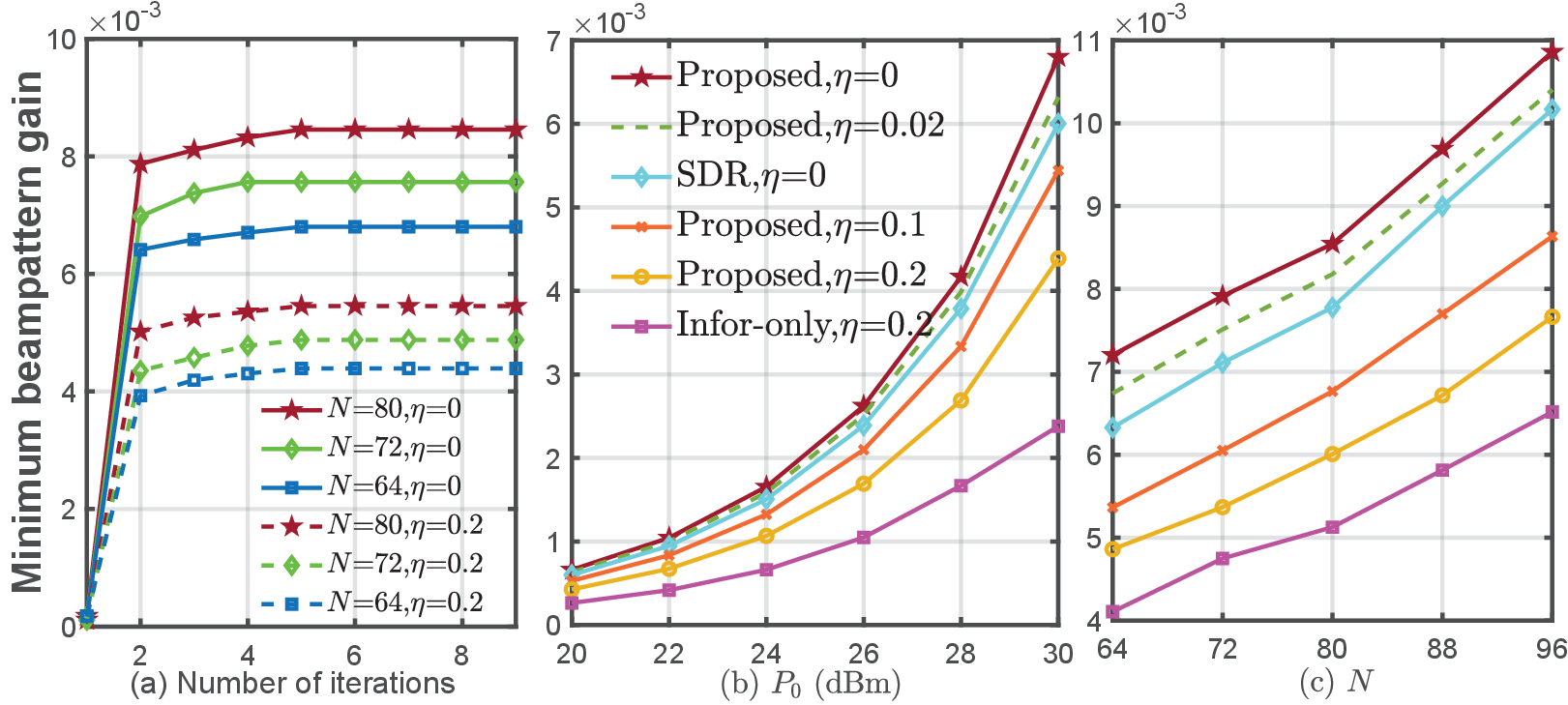} 
    \end{subfigure}
    \caption{(a) The convergence performance of Algorithm 1; (b) The achieved minimum sensing beampattern gain versus the transmit power $P_0$; (c) The achieved  minimum sensing beampattern gain versus BS antenna number $N$.} \label{fig2}
\end{figure}

Fig.~\ref{fig2}(a) shows the convergence performance of the proposed SROCR-based Algorithm~1. The curves for perfect CSI scenarios converge rapidly within about 5 iterations. Under imperfect CSI conditions, Algorithm~1 still achieves convergence within a few iterations, with improved performance as the number of BS antennas increases. Fig.~\ref{fig2}(b) shows the beampattern gain versus the transmit power $P_0$. The proposed scheme shows an improved performance with increasing $P_0$. The proposed scheme outperforms the SDR scheme and the information-only scheme (i.e., the BS only transmits information-bearing signals). This is because the SDR scheme cannot always guarantee a rank-one solution. For information-only scheme, the proposed scheme highlights the importance of joint dual-waveform design in ISAC systems, the incorporation of additional radar sensing signals facilitates enhanced DoFs for ISAC systems. Fig.~\ref{fig2}(c) shows the beampattern gain versus the number of antennas $N$. The performance increases with increasing $N$, the proposed scheme maintains satisfactory performance even under $\eta=0.2$. This demonstrates the strong robustness of the proposed SROCR-based design solution.

\begin{figure} 
    \centering
    \begin{subfigure}[b]{0.81\linewidth} 
        \centering
        \includegraphics[width=\textwidth]{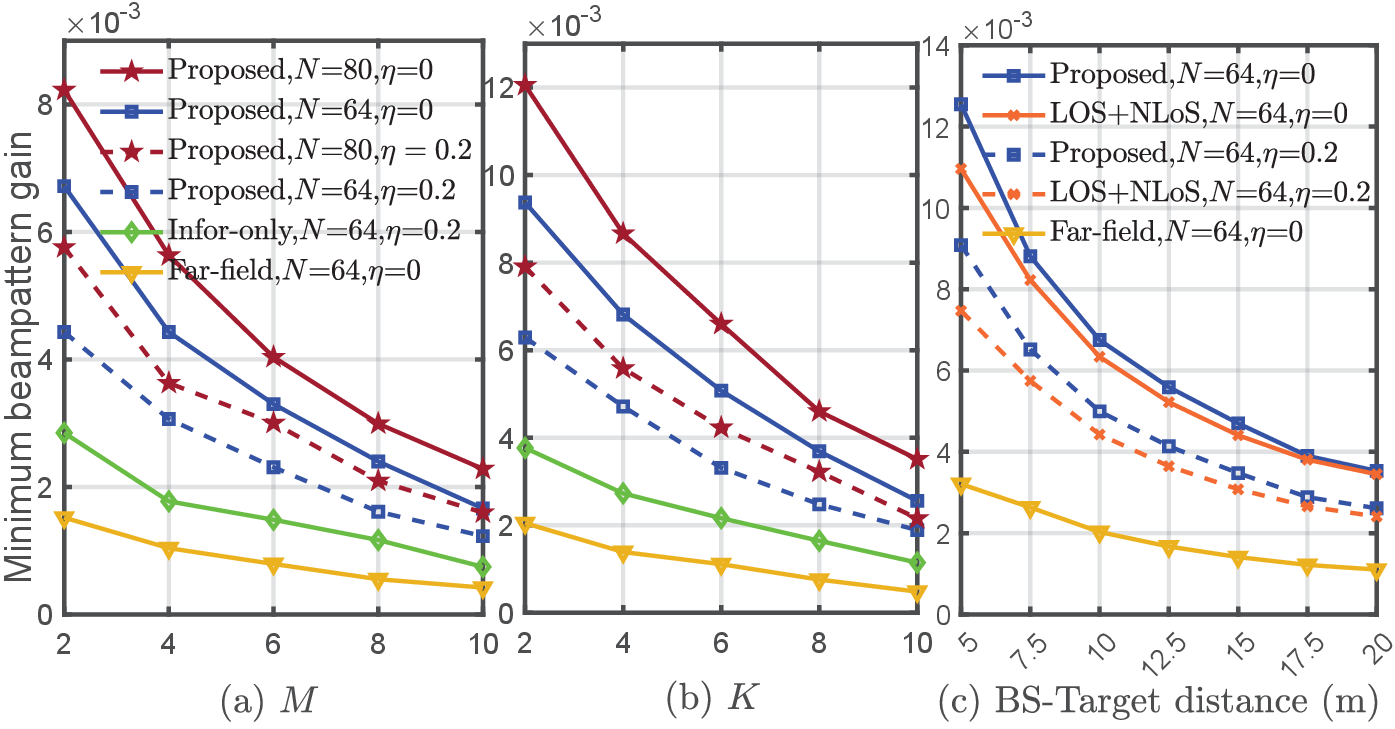} 
    \end{subfigure}
    \caption{ The  achieved minimum sensing beampattern gain; (a) The number of targets $M$; (b) The number of CUs $K$; (c) BS-Target distance.}\label{fig3}
\end{figure}

Figs.~\ref{fig3}(a) and \ref{fig3}(b) show the beampattern gain versus the number $M$ of targets and the number $K$ of CUs, respectively. It is observed that the performance of all the schemes decreases with an increasing number of targets/CUs, attributed to the system requiring a higher transmit power to accommodate multiple targets/CUs. The scheme under the far-field channel model performs inferiorly to those under the near-field channel model with/without CSI errors. Fig.~\ref{fig3}(c) shows the beampattern gain versus the BS-target distance. The performance of all schemes improves as the targets move closer to the BS. The proposed scheme under the LoS/NLoS scenario performs inferiorly to that under the LoS scenario. This is because the NLoS components are not ignored in the channel model \eqref{eq.LoS} and have to be treated as interference. In this case, NLoS components cause the CUs' SINR reduction and energy leakage in the target directions, thereby degrading sensing beampattern gain under a given BS's transmit power budget. It is expected the performance under LoS/NLoS scenarios can be improved by incorporating NLoS components in \eqref{eq.LoS}.

\begin{figure} 
    \centering
    \begin{subfigure}[b]{0.81\linewidth} 
        \centering
        \includegraphics[width=\textwidth]{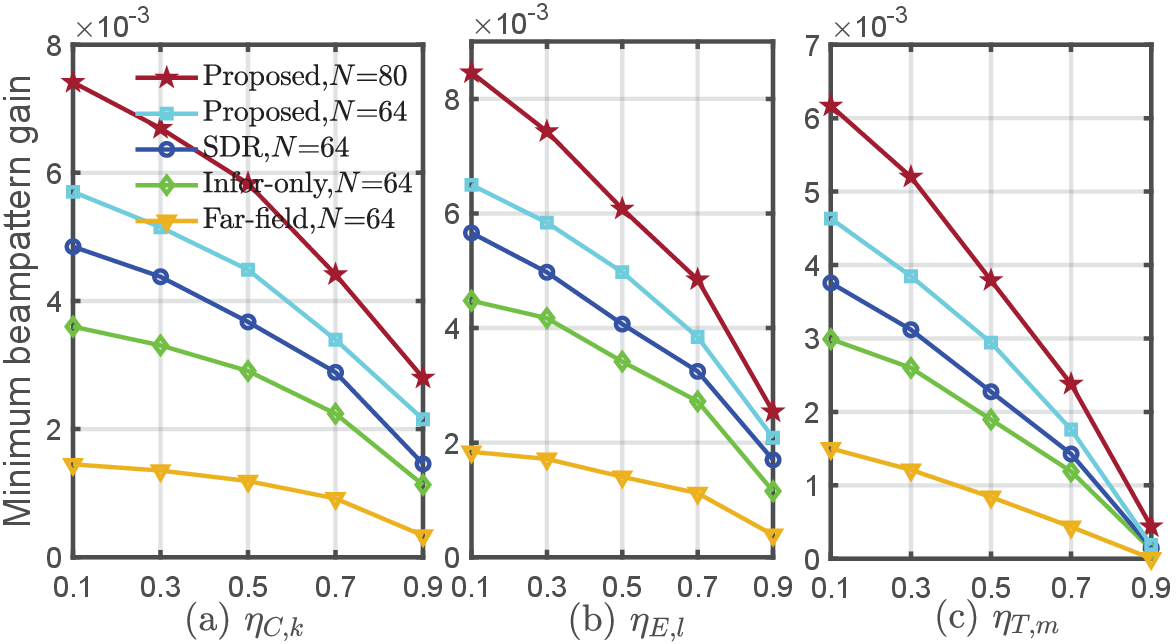} 
    \end{subfigure}
    \caption{The achieved minimum sensing beampattern gain versus the normalized CSI error; (a) $\eta_{E,l}=0$, $\eta_{T,m}=0$; (b) $\eta_{C,k}=0$, $\eta_{T,m}=0$ ; (c) $\eta_{C,k}=0$, $\eta_{E,l}=0$.} \label{fig4}
\end{figure}

 Figs.~\ref{fig4}(a) and \ref{fig4}(b) show the beampattern gain versus $\eta_{C,k}$ and $\eta_{C,k}$, respectively, where all error terms are fixed at zero except for the variation of the X-axis variable. The performance shows a monotonic decrease with increasing estimation error. The proposed scheme improves worst-case beampattern gain through an optimized design even with $\eta_{C,k}=0.9 $ and $\eta_{E,l}=0.9 $. Fig.~\ref{fig4}(c) shows the beampattern gain versus $\eta_{T,m}$. The performance shows rapid deterioration with increasing $\eta_{T,m}$, asymptotically approaching zero. This is because $\eta_{T,m}$ fundamentally affects the accuracy of the beamforming toward the target, which critically governs the performance.

\section{Conclusion} 
In this letter, we proposed a robust beamforming framework for near-field secure ISAC systems to address imperfect CSI challenges. The proposed design solution guarantees the minimum CU's SINR requirement while suppressing information leakage and optimizing the worst-case multi-target sensing beampattern gain. By leveraging the S-Procedure and SROCR agorithm, we developed a solution to the formulated rank-one constrained ISAC design problem with semi-infinitely many constraints. Numerical results demonstrated enhanced security, CSI error robustness, and beamforming precision superiority over the baseline scheme. Note that it is important to extend this work into dynamic scenarios with CUs/targets mobility, where dynamic programming (DP) problem formulation and online reinforcement learning (RL) based solutions are desirable to sequentially optimize robust near-field ISAC beamforming over multiple slots.

\bibliographystyle{IEEEtran}

\end{document}